\documentclass[pre,aps,twocolumn,showpacs]{revtex4}

\usepackage{bm}
\usepackage[]{graphicx}
\usepackage{amsmath}
\usepackage{amssymb}
\usepackage{color}

\begin{document}

\title{Hydrodynamic Irreversibility in Particle Suspensions\\ with Non-Uniform Strain}

\author{Jeffrey S. Guasto$^{1}$, Andrew S. Ross$^{1}$, and J.P. Gollub$^{1,2}$}
\affiliation{$^{1}$Department of Physics, Haverford College, Haverford, Pennsylvania 19041, USA}
\affiliation{$^{2}$Department of Physics, University of Pennsylvania,\\ Philadelphia, Pennsylvania 19104, USA}

\date{\today}

\begin{abstract}
A dynamical phase transition from reversible to irreversible behavior occurs when particle suspensions are subjected to uniform oscillatory shear, even in the Stokes flow limit. We consider a more general situation with non-uniform strain (e.g. oscillatory channel flow), which is observed to exhibit markedly different dynamics.  Self-organization and shear-induced migration only partially explain the delayed, simultaneous onset of irreversibility across the channel. The onset of irreversibility is accompanied by long-range correlated particle motion. This motion leads to particle activity even at the channel center, where the strain is negligible, and prevents the system from evolving into a reversible state.
\end{abstract}

\pacs{47.57.E-, 47.15.G-, 47.52.+j, 83.10.Pp} % Suspensions, Low Re Flow, Chaos in fluid dynamics, Rheology-particle  dynamics

\maketitle

\section{Introduction}

Suspensions of non-Brownian particles were recently shown to flow irreversibly when subjected to sufficiently high, uniform oscillatory strain \cite{Pine2005} even at low Reynolds number (Stokes flows). The significance of this phenomenon is enhanced by the fact that time reversibility is one of the most basic symmetries of physics, and Stokes flows have long been used to demonstrate hydrodynamic reversibility. Furthermore, a distinct threshold strain amplitude was found above which particles fail to return to their original positions upon flow reversal. This system provides one of the few confirmed examples of a non-equilibrium phase transition \cite{Corte2008}. Since suspensions are an important class of fluids in chemical, biological, and industrial applications, the appearance of complex dynamics in such flows is significant. These  experiments have led to a number of theoretical works aimed at explaining the non-equilibrium phase transition \citep{During2009, Menon2009, Santamaria-Holek2009}. 

Random, irreversible particle interactions provide a mechanism for particle self-organization, which leads to a divergent relaxation time at the onset of irreversibility.  Similarities have been demonstrated between this transition and plastic depinning phenomena \cite{Reichhardt2009}. Uniformly sheared particle suspensions can also exhibit self-organized criticality (SOC) \cite{Corte2009} when random interactions and gravitational settling compete.

It has not yet been determined how irreversible particle motion develops in suspension flows that are strained inhomogeneously, a far more general situation. In this paper, we consider the onset of irreversible behavior for dense suspensions (40\% by volume) in an oscillatory channel flow. Here, the strain is largest near the walls and negligible at the center, so some regions may locally exceed the critical strain amplitude, while others do not. We find that the transition to irreversibility is markedly different compared to the Couette flow case studied previously. Self-organization and shear-induced migration contribute to a delayed, simultaneous onset of critical behavior across the channel. However, non-local particle activity generated in high strain regions near the channel walls prevents the system from becoming reversible even where the strain is negligible, i.e. at the channel centerline. In contrast to the Couette flow case, the fluctuation relaxation time does not diverge at the transition.  These experiments provide important guidance for modeling and suggest additional experimental investigations.

\section{Experimental Methods}

\begin{figure}[t]
\begin{center}
\includegraphics[keepaspectratio,width=8.5 cm]{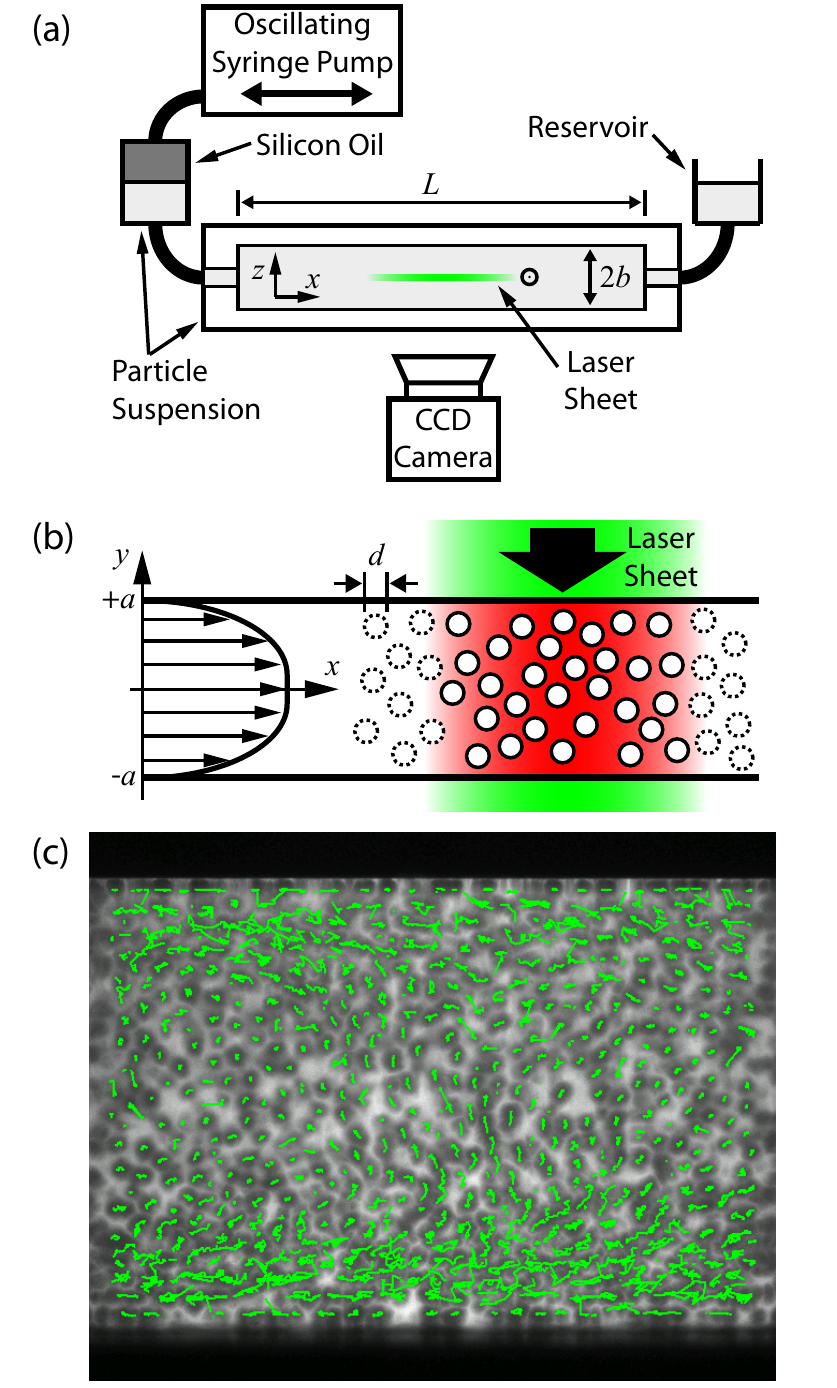}
\end{center}
\caption{(color online) (a) The apparatus uses a syringe pump to drive an oscillating suspension flow. A laser sheet probes a thin plane, which is imaged from below by a CCD camera with a longpass filter. (b) The laser sheet excites fluorescent dye in the interstices between particles, making them visible. (c) Particle tracks overlaid on a multiple exposure image sampled once per period for 35 cycles demonstrates the irreversible behavior especially in high strain regions near the wall.}
\label{fig:1}
\end{figure}

Polymethylmethacrylate (PMMA) particles with a diameter $d = 220 \pm 10$ $\mu$m are suspended at a bulk average volume fraction $\bar{\phi}=0.4$ in an index of refraction and density matched solvent with a viscosity $\mu \approx 3000$ cP \cite{Krishnan1996, Breedveld1998,Corte2008}. The large particle size and solvent viscosity render the particles essentially non-Brownian. A small amount of fluorescent dye (Rhodamine 6G) is added to the fluid for visualization purposes. The channel consists of a slot (width $2a=0.5$ cm, depth $2b=1.5$ cm and length $L=15.0$ cm) machined into an optically clear, cast acrylic block. The walls are polished to remove tool marks and an acrylic plate is bonded to the block to close the channel. The suspension is driven in an oscillatory manner using a syringe pump with a low viscosity silicon oil.  The interface between the oil and suspension occurs well outside the channel test section [Fig. \ref{fig:1}(a)].

A laser sheet (514 nm argon ion laser) is used to probe a thin slice ($\approx 100$ $\mu$m) in the bulk of the suspension. The interrogation region ($\approx 1$ cm long) is centered along the length of the channel far from the ends at mid-depth. A CCD camera with a longpass filter ($>550$ nm) images the fluorescent dye in the interstices of the suspension \cite{Tsai2004,Corte2008} allowing the particles to be tracked [Fig. \ref{fig:1}(b)]. The pump is driven in an approximate square wave pattern (see Video1 \cite{EPAPS2009}) with a peak flow rate of 1 ml/min corresponding to a maximum Reynolds number of $Re < 10^{-3}$, which ensures Stokes flow conditions. Prior to each experiment, the suspension is mixed thoroughly with a magnetic stirrer bar to provide consistent initial conditions.

\begin{figure}[t]
\centering
\includegraphics[keepaspectratio,width=8.4 cm]{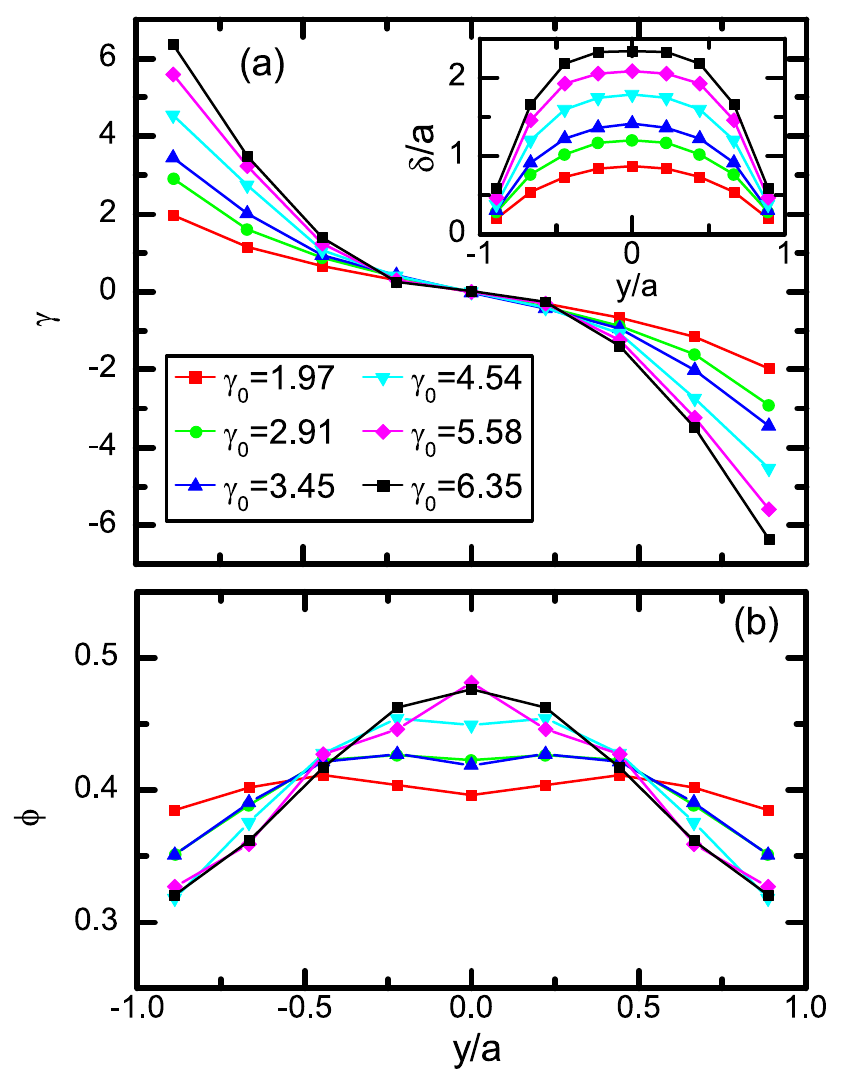}
\caption{(color online) (a) Local strain amplitude estimated from the measured particle displacement profiles (inset). (b) The steady-state particle concentration profiles show significant shear-induced migration at large wall strain amplitudes.}
\label{fig:2}
\end{figure}

For oscillatory motion, the strain amplitude, which governs the mobility of the particles, is varied by adjusting the volumetric displacement amplitude of the reciprocating pump. The suspension is imaged once per period for 500 cycles to capture net deviations from reversible particle trajectories measured using particle tracking algorithms \citep{Ouellette2006}. Afterwards, higher speed imaging (5 fps) provides the steady-state concentration profile, $\phi(y)$, and the measured displacement profile, $\delta(y)$, is used to obtain the strain amplitude, $\gamma=d \delta / dy$ [Fig. \ref{fig:2}]. (All measurements are averaged about the channel centerline to improve the statistical sampling.) The overall strain is parameterized by the largest strain in the system, $\gamma_0$, which occurs at the wall. 

\section{Results}

Irreversible phenomena such as shear-induced migration have long been observed in suspensions \cite{Leighton1986, Leighton1987, Stickel2005}. Numerous experiments and simulations have been aimed at understanding particle migration in pressure-driven pipe and channel flows \cite{Hampton1997, Lyon1998, Nott1994}. Two regimes of migration are observed in oscillatory pressure-driven systems \cite{Butler1999,Morris2001,Yapici2009}, where large oscillation amplitudes cause migration to the centerline, and small amplitudes produce a slight migration toward the wall. In our study, significant migration toward the centerline is observed at large oscillation amplitude, with negligible migration for the smallest strain amplitudes tested [Fig. \ref{fig:2}(b)]. While the bulk effects of these phenomena are well documented, their consequences for particle self-organization and the onset of irreversibility are unknown. 

\subsection{Steady-state Fluctuations}

\begin{figure}[t]
\centering
\includegraphics[keepaspectratio,width=8.4 cm]{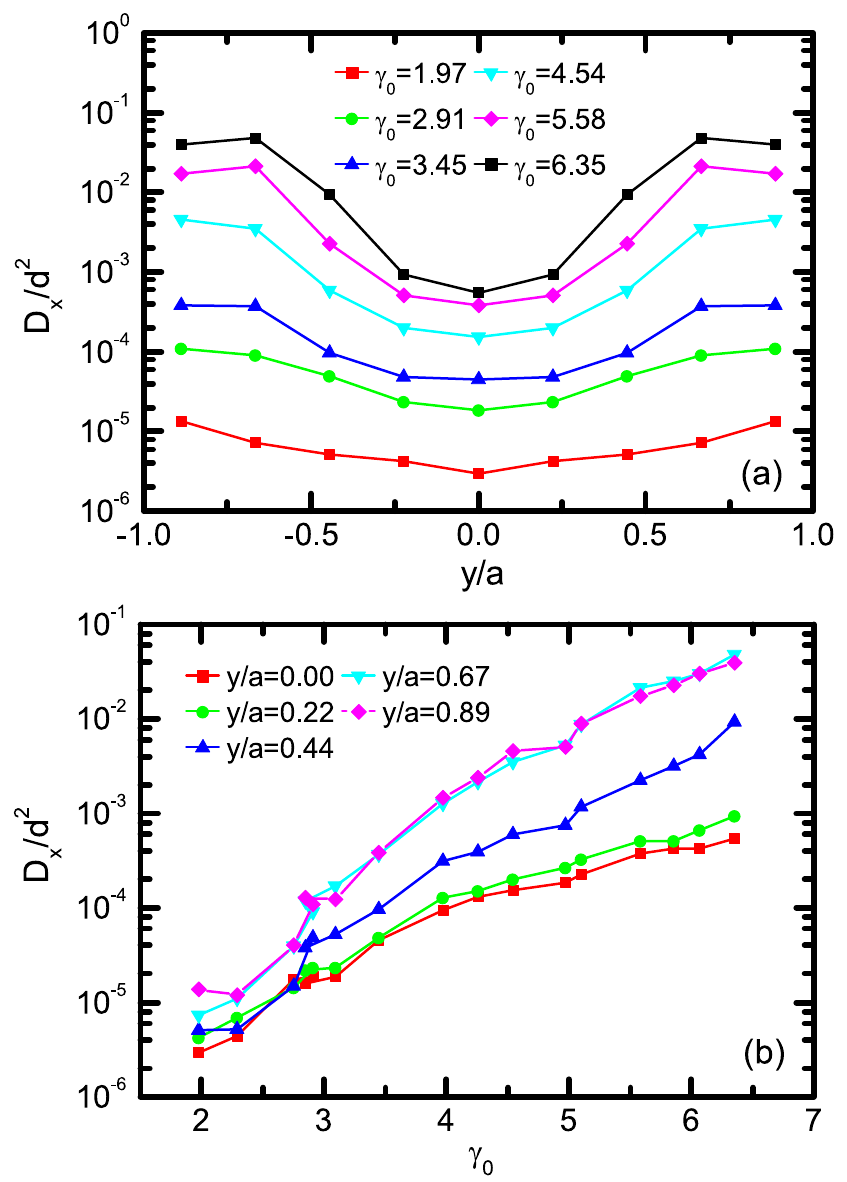}
\caption{(color online) (a) For fixed wall strain amplitude, the particle diffusivity $D_x$ is largest near the channel walls, and also rises with the wall strain amplitude for any fixed position. (b) A subset of the same data shows that the development of irreversibility is simultaneous throughout the channel.}
\label{fig:3}
\end{figure}

The mean square particle displacements (MSDs) measure the degree of irreversibility or particle ``activity''. We use the stream-wise diffusivity, $D_x=\langle\Delta x^2\rangle/2N$, (MSD per cycle), and we omit transients occurring in the first $\sim 200$ cycles when considering steady-state measurements. For small wall strain amplitude, the particle trajectory deviations are negligible and thus reversible. As the wall strain increases, we observe a rapid rise of the particle diffusivity in the high strain regions near the walls, and a markedly slower increase near the center ($D_y$ is somewhat smaller, but shows similar trends) [Fig. \ref{fig:3}(a)]. The irreversible nature of the system at large wall strain amplitude is illustrated in Fig. \ref{fig:1}(c) by periodically sampled particle tracks over 35 cycles (see also Video2 \cite{EPAPS2009}). We note that the largest fluctuations occur slightly away from the channel walls, as previously observed in steady flows \cite{Lyon1998}.

Irreversible behavior does not occur only where the local strain amplitude surpasses a critical value $\gamma \geq \gamma^{c}$  \citep{Pine2005, Corte2008}. As the wall strain amplitude is increased, regions in the center of the channel ($y/a=0$), where the local strain is negligible, show an increase in diffusivity of two orders of magnitude [Fig. \ref{fig:3}(a)]. Although the rate of increase is greater near the wall, the local diffusivity shows a nearly exponential increase with the wall strain amplitude including the centerline, where the local strain amplitude is negligible [Fig. \ref{fig:3}(b)].

\subsection{Transition to Irreversibility}

\begin{figure}[t]
\centering
\includegraphics[keepaspectratio,width=8.4 cm]{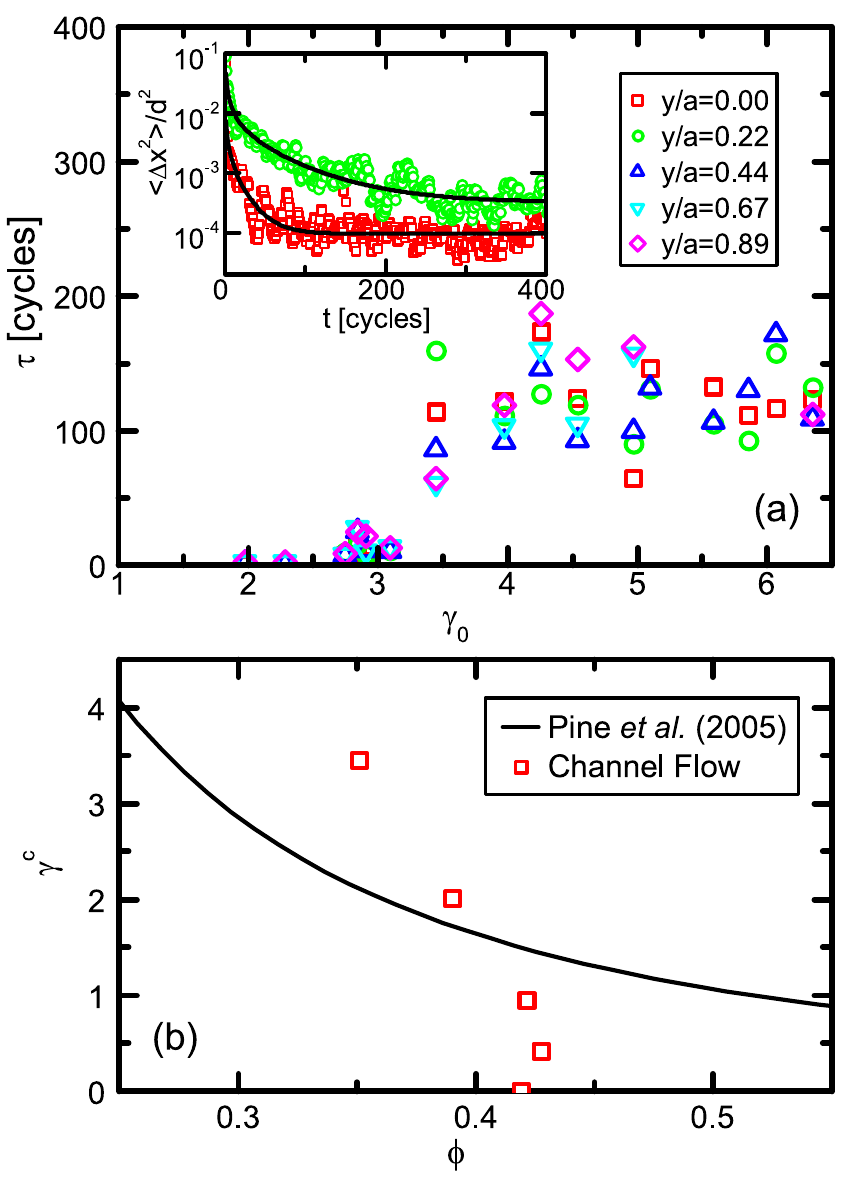}
\caption{(color online) (a) Particle fluctuation relaxation time increases abruptly with wall strain amplitude across the entire channel. Inset: Sample mean square displacement time series for $\gamma_0=2.8$ (\textcolor{red}{$\square$}) and $\gamma_0=4.0$ (\textcolor{green}{$\circ$}) at $y/a=0.44$ with fits. (b) The local critical strain, $\gamma^c$, observed at the global onset $\gamma_0^c \approx 3.4$, as a function of the local concentration, $\phi$. The much steeper variation here compared with the uniformly strained case demonstrates that concentration variations alone are insufficient to account for the simultaneous criticality across the channel.}
\label{fig:4}
\end{figure}

The transition to irreversibility in uniformly sheared particle suspensions has been shown to exhibit the characteristics of a continuous phase transition. The characteristic relaxation time $\tau$ for periodically sampled particle fluctuations to reach a steady state was shown to diverge at the onset of irreversibility in the case of uniformly strained suspensions \cite{Corte2008}. Following the analysis of Cort\'{e} \textit{et al.}, the relaxation time is obtained by fitting a time series of the local particle MSDs consisting of 500 cycles from periodically sampled data to $f\left(t\right)=\left(f^0-f^{\infty}\right) e^{-t/\tau}/t^{\alpha} + f^{\infty}$, where $\alpha=0.6$ and $f^0$ and $f^{\infty}$ are the initial and steady values of the particle MSD, respectively \cite{Corte2008}. Figure \ref{fig:4}(a) (inset) shows sample time series for both low and high strain with fits for $y/a=0.44$ (see also Video2 \citep{EPAPS2009}). 

For small wall strains, particles reorganize quickly within a few cycles ($\tau<10$ cycles) and settle into a reversible steady state. As the wall strain is increased, $\tau$ increases abruptly across the entire channel at $\gamma^c_0 \approx 3.4$ indicating a simultaneous onset of irreversible behavior [Fig. \ref{fig:4}(a)]. Thus, the local critical strain is elevated at the wall and depressed near the channel center (compared to $\gamma^c = 1.6$ for $\bar{\phi}=0.4$ in uniform shear) \cite{Pine2005}. We also note that $\tau$ does not diverge, but rather maintains an elevated value ($\approx 130$ cycles) as $\gamma_0$ increases further. 

The transition to irreversibility is sensitive to the particle concentration \cite{Pine2005}, and for the conditions tested here, the volume fraction can vary from 30\% at the walls to 50\% in the center of the channel as a result of shear-induced migration [Fig. \ref{fig:2}(c)]. The delayed onset of irreversibility can be explained \textit{partially} by this effect. In Fig. \ref{fig:4}(b), we examine the variation in local critical strain amplitude as a function of the local particle concentration (for the single detected critical wall strain, $\gamma_0^c$) and compare it to the critical strain amplitude for the previously studied Couette flow. The much steeper variation here compared with the uniformly strained case shows that concentration variations alone are insufficient to explain the simultaneous criticality across the channel.

\subsection{Long-range Correlated Fluctuations}

\begin{figure}[t]
\centering
\includegraphics[keepaspectratio,width=8.4 cm]{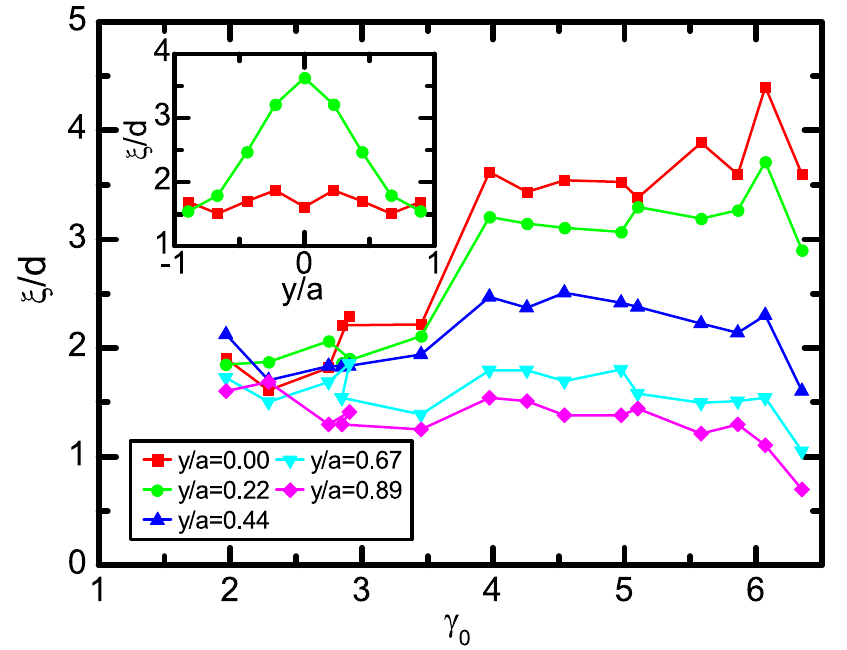}
\caption{(color online) Particle motions in the center of the channel become correlated over a longer range above the critical wall strain amplitude, which indicates that the fluctuations do not arise from local interactions. Inset: Sample correlation length profiles for $\gamma_0=2.3$ (\textcolor{red}{$\square$}) and $\gamma_0=4.0$ (\textcolor{green}{$\circ$}).}
\label{fig:5}
\end{figure}

An important additional factor is the role of irreversible activity near the walls, which appears to influence the core of the channel some distance away.  As evidence for this effect, we note that coordinated particle motion is observed in the center of the channel once  $\gamma_0^c$ has been exceeded  [Fig. \ref{fig:1}(c)] (see also Video2 \citep{EPAPS2009}). Other works have drawn attention to the importance of correlated particle motion or ``transient clusters'' to the rheology and flow of suspensions \cite{Mills1995}.  We quantified this collective motion by a displacement correlation function $g(r^\prime) = \langle \vec{\Delta r_i} \cdot \vec{\Delta r_j}\rangle / \langle \vec{\Delta r_i} \cdot \vec{\Delta r_i}\rangle$, where $r^\prime=|\vec{r_j}-\vec{r_i}|$ is the separation between any two particles. The corresponding correlation length is denoted by $\xi$; the largest measurable correlations are limited by the finite system size. Below the critical strain,  particle motion in the periodically sampled data is minimal and spatially uncorrelated across the channel [Fig. \ref{fig:5} (inset)]. However, once the critical strain is surpassed, the correlation length increases in the center, but remains small near the walls [Fig. \ref{fig:5}]. The larger correlation length of the particle motion in the center region implies that these fluctuations originate from irreversible disturbances some distance away rather than from local particle interactions.

\section{Conclusion}

We have studied the onset of irreversible motion in particle suspensions in an oscillatory channel flow, where the shear strain is non-uniform.  The transition is shown to be quite different from the previously studied Couette flow case in that it does not exhibit a divergent relaxation time.  Strikingly, the transition is delayed to higher wall strain amplitude, and occurs simultaneously across the channel, even where the local strain is negligible.  We find that self-organization and shear-induced migration alone cannot explain the simultaneous onset.  The transition to irreversible dynamics is not governed purely by the local strain, but depends also on disturbances generated some distance away.  This work is an important step in understanding the coupled behavior of irreversible particle motion (activity), local particle concentration, and imposed strain \cite{Menon2009}.  However, many challenges remain including consideration of nonlocal effects in models of the transition, the possible role of finite system size, and other materials such as polydisperse or soft suspensions.

\section*{Acknowledgement}
We thank David Pine and Laurent Cort\'{e} for helpful discussions, and Bruce Boyes for technical assistance. This work was supported by NSF Grant DMR-0803153.

\end{document}